# Transverse relativistic effects in paraxial wave interference

Konstantin Y. Bliokh[1,2], Yana V. Izdebskaya[3,4], and Franco Nori[1,5]

[1]Advanced Science Institute, RIKEN, Wako-shi, Saitama 351-0198, Japan
[2]A. Usikov Institute of Radiophysics and Electronics, NASU, Kharkov 61085, Ukraine
[3]Nonlinear Physics Center, Research School of Physics and Engineering,
The Australian National University, Canberra ACT 0200, Australia
[4]Department of Physics, Taurida National University, Simferopol 95007, Ukraine
[5]Physics Department, University of Michigan, Ann Arbor, Michigan 48109-1040, USA

**Abstract.** We consider relativistic deformations of interfering paraxial waves moving in the transverse direction. Owing to *superluminal* transverse phase velocities, noticeable deformations of the interference patterns arise when the waves move with respect to each other with *non-relativistic* velocities. Similar distortions also appear on a mutual tilt of the interfering waves, which causes a phase delay analogous to the relativistic time delay. We illustrate these observations by the interference between a vortex wave beam and a plane wave, which exhibits a pronounced deformation of the radial fringes into a fork-like pattern (relativistic Hall effect). Furthermore, we describe an additional relativistic motion of the interference fringes (a counter-rotation in the vortex case), which becomes noticeable at the same non-relativistic velocities.

## 1. Introduction

Special relativity owes its origin to Maxwell wave equations and optical interference experiments with moving sources or observers [1]. Among its most fundamental consequences are the relativity of time and impossibility of information transfer faster than the speed of light *c*. Nonetheless, during the past century, physicists continuously pursued the speed-of-light limit and offered various examples of superluminal wave motion.

It should be noticed that the *phase* velocity of a wave is not related to the information transfer and, hence, is not limited by $c$. Indeed, consider an optical plane wave propagating at an angle $\pi/2 - \theta$ to the *x*-axis. One can easily see that the wave fronts move with velocity $u_{\text{ph}\,x} = c/\sin\theta > c$ along the *x*-axis, which can be arbitrarily large when $\theta \to 0$. Unlimited superluminal motion can also occur for any phase structures (e.g., dislocations) and interference fringes [2,3]: as Berry pointed out, these are "forms and not things, and so cannot be used as signals" [2].

In addition, there are numerous examples of superluminal *group* velocities, which appear in dispersive media [4], special X-shaped solutions of wave equations [5], tunneling of wavepackets (Hartman paradox) [6], and, locally, in evanescent waves or complex wave superpositions [7]. However, in all these cases, the group velocity again transfers "forms", whereas the *signal* velocity never exceeds *c* [4,6,8].

In contrast with the studies [4−8], which tried to construct objects moving faster than *c*, the purpose of our work is quite opposite. Namely, we are wondering if strong relativistic effects can appear for an observer moving *much slower* than *c*. For mechanical bodies (i.e., "things"), the relativistic deformations become noticeable when the observer motion approaches the speed of light: $v \sim c$ [1]. However, below we show that, owing to the superluminal phase velocities of waves, $u_{\text{ph}} \gg c$, the pronounced relativistic distortions of the interference "forms" might appear for the observer's velocities $v \ll c$. This offers a new avenue for visualizations and experimental tests of special relativity using paraxial wave interference and *non-relativistic* transverse motion.

## 2. Relativistic deformations: Lorentz contraction and velocity addition

Special relativity is based on the Lorentz transformations of space-time, which describe transitions from a 'laboratory' reference frame to a frame moving with velocity $\mathbf{v}$:

$$t' = \gamma\left(t - \boxed{\mathbf{v}\cdot\mathbf{r}/c^2}\right), \quad \mathbf{r}' = \gamma(\mathbf{r} - \mathbf{v}t). \tag{1}$$

Here $\gamma = 1/\sqrt{1-v^2/c^2}$ is the Lorentz factor, and quantities in the moving frame are indicated by primes. Importantly, the $\mathbf{r}$-dependent time delay, shown inside the red box in Eq. (1), revises the concept of simultaneity and causes interesting distortions of objects when observed in a moving reference frame [1]. Such distortions are absent in non-relativistic physics based on Galilean transformations and the invariance of time. As we argue below, one can distinguish two types of relativistic deformations:
(i) Lorentz length contraction of motionless objects;
(ii) Shape distortions of moving objects, related to the relativistic velocity addition.

Throughout this paper we consider reference frames moving with respect to each other with velocity $v$ in the $x$-direction. Let a material point move with velocity $u$ along the $x$-axis of the laboratory frame: $x(t) = x_0 + ut$. Then, applying the Lorentz transformation (1), one can find that its coordinate in the moving frame becomes

$$x'(t') = \frac{\boxed{\gamma^{-1}x_0} + (u-v)t'}{\boxed{1 - uv/c^2}}. \tag{2}$$

For a motionless point, $u=0$, the coordinate $x'(0) = \gamma^{-1}x_0$ indicates the *Lorentz contraction*, whereas for moving point the velocity $u' \equiv dx'/dt' = (u-v)/(1-uv/c^2)$ yields the relativistic *velocity addition* formula. In the general case $u \neq 0$, the coordinate $x'(0)$ in Eq. (2) indicates the transformation of the $x$-scale of the object and includes both the Lorentz contraction [shown inside the green box in (2)] and the velocity addition effect [shown inside the orange box in (2)]. Although both of these distortions originate from the same Lorentz transformation of time (1), below we show that they can occur independently in various situations. Note that the velocity-addition deformation is a first-order effect in $v/c$ and also depends on $u/c$, whereas the Lorentz contraction is a second-order effect $\sim v^2/c^2$.

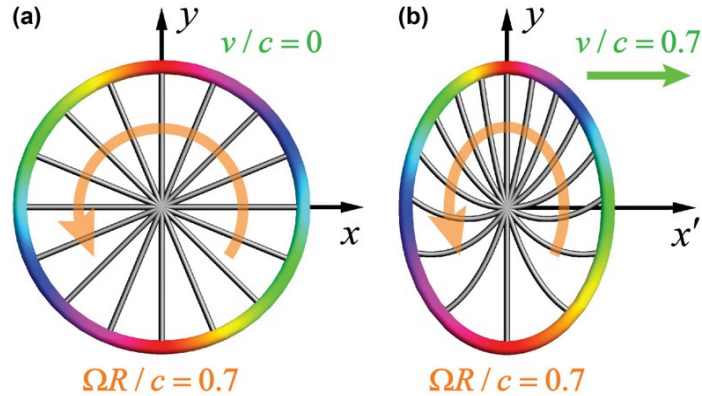

**Figure. 1.** Virtual snapshots of a relativistic flywheel of radius $R$ rotating with angular velocity $\Omega$: **(a)** in the laboratory frame and **(b)** in the frame moving with velocity $v$ in the $x$-direction. Two relativistic deformations are seen in **(b)**: (i) the Lorentz $x$-contraction squeezing the circle into an ellipse (becomes noticeable at $v \sim c$) and (ii) a characteristic distortion of the radial spokes along the orthogonal $y$-direction (relativistic Hall effect) [9,10]. The latter effect is caused by the relativistic addition of the rotational velocity $u = \Omega R$ and frame velocity $v$, so that it becomes noticeable at $\Omega R v \sim c^2$.

A nice illustration of the above relativistic deformations appears when observing a spinning body (flywheel) in a moving reference frame [9], as shown in Fig. 1. First, the circular flywheel experiences the Lorentz $x$-contraction with the factor of $\gamma^{-1}$, and becomes elliptical. Second, the rotating radial spokes in the wheel become distorted and asymmetrically redistributed along the orthogonal $y$-axis because of opposite velocity additions on the $y > 0$ and $y < 0$ sides of the wheel. The Lorentz contraction depends only on the frame motion and becomes noticeable at $v \sim c$. In contrast, the *y*-deformation is essentially related to the rotational velocity of the wheel, $u = \Omega R$ ($R$ and $\Omega$ are the radius and angular velocity of the wheel, respectively); this deformation becomes noticeable at $uv \sim c^2$. Noteworthily, the *y*-distortion of a spinning body on a Lorentz boost in the *x*-direction is intimately related to the Lorentz transformation of the angular momentum, and can be regarded as the relativistic Hall effect [10].

## 3. Deformations of wave intensity and phase: Superluminal wavefronts

Recently, we found [10] that "spinning" waves carrying angular momentum, the so-called *vortex beams*, also experience relativistic deformations resembling those in Fig. 1. The vortex beams are well known and widely used in optics [11]. Few years ago they were also described for quantum electrons [12] and generated experimentally in electron microscopes [13].

Let us consider a scalar monochromatic vortex beam propagating along the *z*-axis. In what follows we are interested in the wave distributions in the transverse $(x, y)$-plane and omit all *z*-dependences. Then, the vortex beam is described by the wave function

$$\psi(x,y,t) \propto A(r)\exp(i\ell\varphi - i\omega t), \quad (3)$$

where $(r, \varphi)$ are the polar coordinates in the $(x, y)$-plane, $\ell = 0, \pm 1, \pm 2, ...$ is the vortex charge (the quantum number of the angular momentum along the *z*-axis), $\omega$ is the frequency, and $A(r)$ is the radial amplitude distribution. Hereafter, we assume the Laguerre-Gaussian beams [11] with the zero radial index and $A(r) \propto (\kappa r)^{|\ell|} \exp\left[-(\kappa r)^2\right]$, where $\kappa = k_\perp$ is the characteristic radial wave number. Such beams have an annular intensity distribution with characteristic radius $R \sim \sqrt{|\ell|}/\kappa$. The phase fronts of the vortex (3) represent $|\ell|$ radial lines which rotate in the $(x, y)$-plane with the angular velocity $\Omega_{ph} = \omega/\ell$. Upon transition to the moving reference frame, the annular intensity distribution $I = |\psi|^2 = |A(r)|^2$ experiences the Lorentz *x*-contraction with the factor $\gamma^{-1}$. At the same time, the rotating radial phase fronts of the vortex undergo relativistic Hall-effect *y*-deformations entirely similar to the spokes of a spinning flywheel in Fig. 1 (see Fig. 2b) [10].

Thus, there is a correspondence between the relativistic deformations of mechanical bodies and waves, but there is also a remarkable difference. Namely, the transverse velocity of the wavefront motion, i.e., the *phase velocity*, is *superluminal*. Indeed, consider a paraxial wave propagating mostly along the *z*-axis, with the longitudinal wave number $k_z \simeq k$ and a characteristic transverse wave number $k_\perp \sim \kappa \ll k$. To quantify the paraxiality, we will use the small parameter $\theta = \kappa/k \ll 1$. Then, the transverse phase velocity in the $(x, y)$ plane is estimated as

$$u_{ph} \sim \frac{\omega}{\kappa} = \frac{c}{\theta} \gg c \quad (4)$$

(for the sake of simplicity we assume waves with $\omega = ck$). For instance, in the above optical-vortex example, the rotational velocity of the wavefronts at the beam radius is $u_{ph} = \Omega_{ph} R \sim c/\theta \gg c$. Therefore, the velocity-addition deformations of the wavefronts become noticeable at

$$v \sim \frac{c^2}{u_{ph}} \sim \theta c \ll c, \quad (5)$$

i.e., at essentially *non-relativistic* velocities of the frame motion. This is demonstrated in Fig. 2, which displays the transverse intensity, current, and phase distributions for the paraxial vortex beam observed in reference frames moving with small velocities $v \sim \theta c$. One can see no Lorentz contraction in the vortex intensity distribution but a pronounced *y*-distortion of the wavefronts due to the relativistic velocity addition with $u_{ph} \sim c/\theta$.

Of course, superluminal motion of the wavefronts is non-observable *per se*. But the shape of the phase fronts plays a crucial role in the wave interference. Then the question arises: *Can one observe relativistic deformations of the wave interference patterns at non-relativistic velocities?* We address this question below.

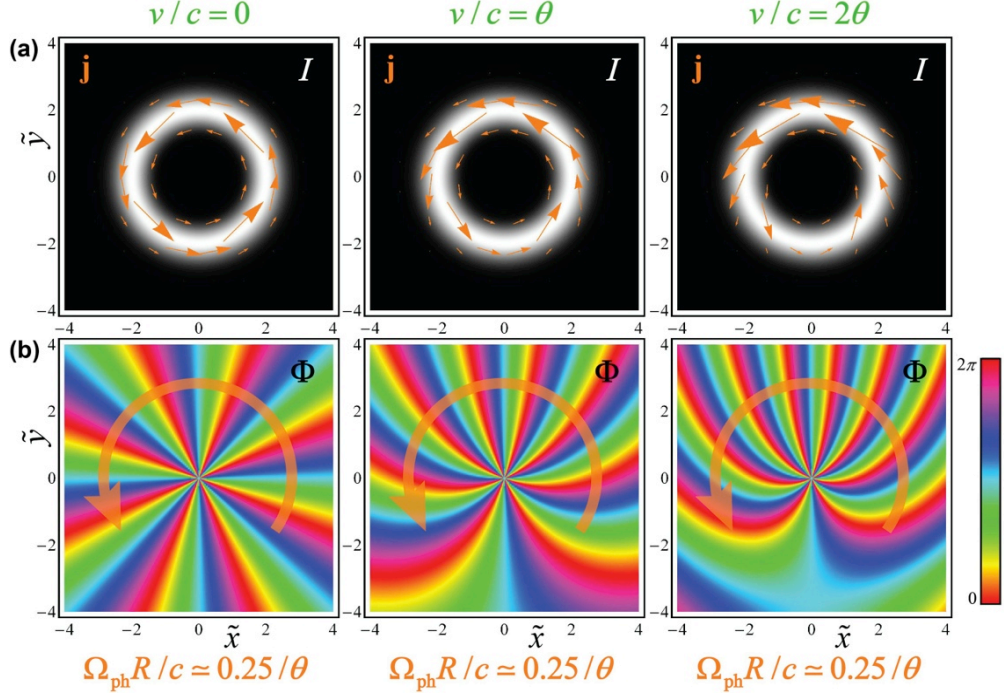

**Figure 2.** Instant transverse distributions of: **(a)** Intensity $I = |\psi|^2$, current $\mathbf{j} = \text{Im}(\psi^* \nabla \psi) = I \nabla \Phi$, and **(b)** phase $\Phi = \arg \psi$ in the paraxial vortex beam (with $\ell = 8$ and paraxiality parameter $\theta = \kappa / k \ll 1$) propagating along the z-axis. The dimensionless coordinates $\tilde{x} = \kappa x'$ and $\tilde{y} = \kappa y$ are used. The distributions are shown in the frames moving in the x-direction with velocities $v / c = 0, \theta, 2\theta \ll 1$. Such non-relativistic velocities make the Lorentz x-contraction of the intensity distribution negligible, but nonetheless drastically deform wavefronts, similar to the Hall-effect y-distortion of the flywheel spokes in Fig. 1. This is explained by the relativistic velocity addition with the superluminal motion of the radial wavefronts (4): $u_{ph} / c = \Omega_{ph} R / c \simeq 0.25 / \theta \gg 1$ ($R \simeq 2\kappa^{-1}$ is the beam radius and $\Omega_{ph} = \omega / 8$ is the angular velocity of the wavefronts rotation).

## 4. Transformations of the wave interference patterns

Let us consider a generic wave interference of two complex scalar fields:

$$\psi(\mathbf{r},t) = \psi_1(\mathbf{r},t) + \psi_2(\mathbf{r},t). \tag{6}$$

The interference pattern is described by the resulting intensity distribution, $I(\mathbf{r},t) = |\psi(\mathbf{r},t)|^2$. Note that interference fringes can move with arbitrarily large velocities, but, of course, can be observed only in the case of subluminal motion.

To find transformations of the interference pattern on transition to the moving reference frame, one should substitute the Lorentz transformation (1) in the wave function (6). One can distinguish two basic types of relativistic effects in wave interference:

(i) The observer moves with respect to both waves and observes the same interference picture but in the moving frame;

(ii) The observer and the second wave move with respect to the first wave. In other words, the second wave is used as a probe attached to the observer and sensing the first wave in the moving frame.

For these two cases, the wave functions in the moving frame can be written, respectively, as

$$\psi'(\mathbf{r}',t') = \psi_1[\mathbf{r}(\mathbf{r}',t'), t(\mathbf{r}',t')] + \psi_2[\mathbf{r}(\mathbf{r}',t'), t(\mathbf{r}',t')], \tag{7a}$$

$$\psi'(\mathbf{r}',t') = \psi_1[\mathbf{r}(\mathbf{r}',t'), t(\mathbf{r}',t')] + \psi_2(\mathbf{r}',t'). \tag{7b}$$

Here $\mathbf{r}(\mathbf{r}',t')$ and $t(\mathbf{r}',t')$ denote the Lorentz transformation given by Eq. (1). In the following Sections 4.1 and 4.2 we analyze the relativistic deformations in the interference patterns $I'(\mathbf{r}',t') = |\psi'(\mathbf{r}',t')|^2$ for the two cases (7a) and (7b), respectively.

## 4.1. *Moving interference patterns*

First, we examine the frame moving with respect to both waves. In this case, Eq. (7a) shows that the interference intensity pattern is transformed as any material object, i.e., via the Lorentz transformations (1): $I'(\mathbf{r}',t') = I[\mathbf{r}(\mathbf{r}',t'), t(\mathbf{r}',t')]$. This is quite natural since in quantum mechanics any matter distribution is associated with the intensity of the wave function.

As the simplest example, let us consider the interference of two plane waves propagating in the $(z,x)$ plane, with transverse wave numbers $k_{x1,2} = \kappa_{1,2}$. As before, we are only interested in the distributions in the $z = 0$ plane and omit all $z$-dependences. Thus, the two interfering wave functions are:

$$\psi_1(x,t) = \exp(i\kappa_1 x - i\omega_1 t), \quad \psi_2(x,t) = \exp(i\kappa_2 x - i\omega_2 t). \tag{8}$$

The interference pattern for these waves, $I(x,t)$, represents an array of fringes with period $\Delta = 2\pi/|\kappa_1 - \kappa_2|$ and moving with velocity $u_f = (\omega_1 - \omega_2)/(\kappa_1 - \kappa_2)$ along the $x$-direction. Let us now choose one fringe in this interference pattern, which has a coordinate $x(t) = x_0 + u_f t$ in the laboratory frame. Then, performing the Lorentz transformation (1) and (7a), one can readily ascertain that the coordinate of this fringe in the moving frame is given by Eq. (2) with $u = u_f$ and the corresponding Lorentz contraction and velocity addition. The only difference is that the fringe motion can be superluminal [3], $u_f \gg c$, and then the relativistic velocity-addition effects formally occur at $v \sim c^2/u_f \ll c$. However, superluminal fringes and, hence, their deformations remain fundamentally unobservable.

It is worth noticing that the Lorentz transformation (1), when applied to a plane wave $\exp(i\mathbf{k}\cdot\mathbf{r} - i\omega t)$, results in the following transformation of the wave parameters:

$$\omega' = \gamma(\omega - \mathbf{k}\cdot\mathbf{v}), \quad \mathbf{k}' = \gamma\left(\mathbf{k} - \boxed{\omega\mathbf{v}/c^2}\right), \tag{9}$$

so that the wave function becomes $\exp(i\mathbf{k}'\cdot\mathbf{r}' - i\omega' t')$ in the moving frame. The shift of the wave vector in Eq. (9) [shown inside the red box] originates from the time delay in Eq. (1), and it is this shift that causes deformations of the interference patterns. In the paraxial geometry $k_z \simeq k$, $v_x = v \ll c$, the transformation (9) represents a *tilt* of the wave vector in the $(z,x)$-plane by the angle $\alpha \simeq v/c$. This will be used in what follows.

For comparison with the examples in Sections 2 and 3, let us consider now an interference pattern which mimics a spinning flywheel in Fig. 1. Such pattern appears when interfering the co-propagating vortex beam (3) and plane wave (the $z$-dependences are omitted) [14]:

$$\psi_1(x,y,t) = A(r)\exp(i\ell\varphi - i\omega_1 t), \quad \psi_2(t) = \exp(-i\omega_2 t). \tag{10}$$

The intensity distribution of the superposition (10) represents a circular array of $|\ell|$ radial "spokes" with vortex radius $R \sim |\ell|/\kappa$, rotating with angular velocity $\Omega_f = (\omega_1 - \omega_2)/\ell$ (Fig. 3a). Thus, the velocity of the circular motion of the radial fringes is $u_f \sim (\omega_1 - \omega_2)/\kappa$, and it can take on any values depending on the frequency difference. Figure 3 shows deformations of the interference pattern of waves (10) in the moving frame. When $u_f \sim c$ and $v \sim c$, both the Lorentz $x$-contraction of the circle and velocity-addition $y$-distortion of the spokes appear (Fig. 3b), entirely similar to those in Fig. 1. At the same time, when $u_f \gg c$ and $v \sim c^2/u_f \ll c$, the Lorentz contraction is negligible, while the velocity-addition deformation of the radial fringes is present (Fig. 3c), akin to the distortion of the vortex wavefronts in Fig. 2b. Still, as we mentioned before, this superluminal effect cannot be detected.

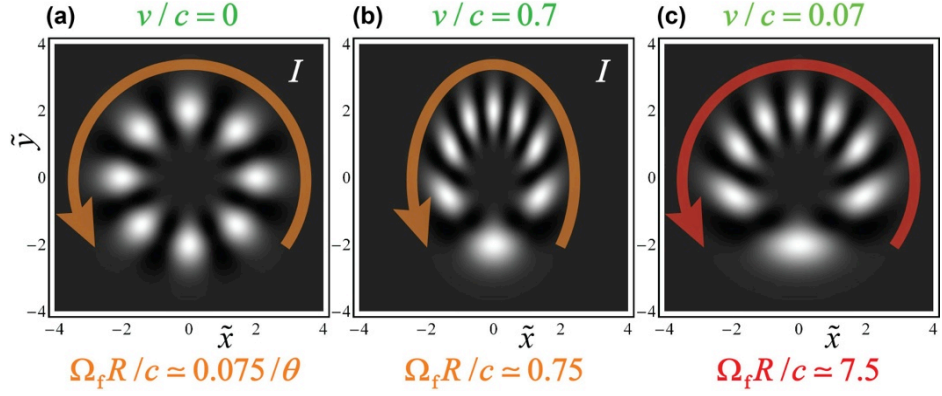

**Figure 3.** Transverse intensity pattern for interference (10) of co-propagating paraxial vortex beam ($\ell = 8$) and plane wave. The dimensionless coordinates $\tilde{x} = \kappa x'$ and $\tilde{y} = \kappa y$ are used. **(a)** In the laboratory frame, the radial fringes rotate with angular velocity $\Omega_f = (\omega_1 - \omega_2)/8$ and linear velocity $u_f \simeq \Omega_f R$ ($R \simeq 2\kappa^{-1}$ being the vortex radius) which can take on arbitrary values depending on the wave parameters. **(b)** In the moving frame with $v \sim c$ and with the fringe velocity $\Omega_f R \sim c$, the pattern shows both the Lorentz $x$-contraction and velocity-addition $y$-distortion of the fringes, entirely similar to the mechanical flywheel in Fig. 1. **(c)** Choosing parameters corresponding to a superluminal fringe velocity $\Omega_f R \gg c$, the velocity-addition distortion (but no Lorentz contraction) occurs for non-relativistic frame motion, $v \ll c$, but cannot be observed.

### 4.2. *Waves moving with respect to each other*

Finally, we examine the second type of relativistic interference, when the two waves move with respect to each other. Assuming that the observer is attached to the second wave, the transformation to the moving frame is described by Eq. (7b). From here on, we consider *non-relativistic* velocities of the frame motion, $v \ll c$, so that $\gamma \simeq 1$ in the main approximation.

Performing the transformation (7b) with (1) in the simplest case of two interfering plane waves (8), we find that the interference fringe with the coordinate $x(t) = x_0 + u_f t$ in the laboratory frame will have the following coordinate in the moving frame:

$$x'(t') \simeq \frac{x_0 + \left(u_f - v\frac{\kappa_1}{\kappa_1 - \kappa_2} + \boxed{\frac{v^2}{2c^2}\frac{\omega_1}{\kappa_1 - \kappa_2}}\right)t'}{\boxed{1 - \frac{u_{ph1} v}{c^2}\frac{\kappa_1}{\kappa_1 - \kappa_2}}}, \tag{11}$$

where $u_{ph1} = \omega_1/\kappa_1$ is the phase velocity of the first wave. The Lorentz contraction is absent in Eq. (11) since $v \ll c$ but the velocity-addition effects are present. The most important difference in the velocity-addition denominator of Eq. (11) as compared to Eq. (2) is that it contains the phase velocity of the *first* wave, and is *independent* of the fringe velocity $u_f$. Owing to this, the velocity-addition distortions can be observed for $u_{ph1} \gg c$ but non-relativistically moving (or even motionless) fringes, $u_f \ll c$. And this is the desired observable relativistic deformation at $v \ll c$, described by the denominator of Eq. (11).

In addition, we kept a second-order term $\sim v^2/c^2$ [shown inside the green box in Eq. (11)], which originates from the $\gamma$-factor increment of the frequency in Eq. (9): $\omega_1' \simeq (1 + v^2/2c^2)\omega_1 - \kappa_1 v$. Such $\gamma$-scaling of the frequency of a moving wave contributes to the transverse relativistic Doppler effect – a frequency counterpart of the Lorentz contraction [1]. Since $\omega_1 = k_1 c = \kappa_1 c/\theta$, we find that, in the paraxial approximation, the relativistic-Doppler term in Eq. (11) can make a noticeable contribution $\sim v$ at non-relativistic velocities (5): $v \sim \theta c$. Furthermore, when $u_f = 0$ and $\kappa_1 = 0$, only this term causes motion of the fringes in the moving frame.

Let us illustrate these results by considering the interference of the co-propagating vortex and plane wave, Eq. (10). We set $\omega_1 = \omega_2 \equiv \omega$ so that the fringes do *not* rotate in the laboratory frame: $u_f = 0$. Figure 4a shows the deformation of the interference fringes upon motion with non-relativistic velocities $v \sim \theta c$, Eq. (5). The characteristic $y$-distortions of the non-rotating radial fringes appear. One can say that they represent deformations of the superluminal vortex wavefronts in Fig. 2b, revealed by the interference with a plane wave in the moving frame. Thus, we conclude that *non-relativistic motion can produce pronounced relativistic deformations of the intensity pattern when the two waves move with respect to each other in the transverse direction*.

Moreover, the $\gamma$-factor scaling of the wave frequency in the moving frame induces *rotation* of the above interference pattern (see Fig. 4a). It is easier to describe this rotation in the reference frame attached to the vortex. In this frame, the vortex and plane-wave frequencies become $\omega_1' = \omega$ and $\omega_2' \simeq (1 + v^2/2c^2)\omega$, which yields the following angular velocity of the rotation of fringes:

$$\Omega_\mathrm{f} \simeq -\frac{v^2}{2c^2}\frac{\omega}{\ell} = -\frac{v}{2c}\frac{\kappa v}{\theta \ell}. \qquad (12)$$

This equation shows that the pattern rotates in the direction *opposite* to the vortex phase-front rotation. Apparently, the counter-rotation of the fringes can be associated with the transverse relativistic Doppler effect, as it is equivalent to the $\gamma^{-1}$ red-shift scaling of the vortex frequency. Furthermore, the linear velocity of the fringe motion becomes noticeable, $u_\mathrm{f} \simeq \Omega_\mathrm{f} R \sim v$, under the same non-relativistic condition (5): $v \sim \theta c$. Thus, *alongside the deformation of fringes caused by the velocity-addition effect, one can observe the rotation of fringes induced by the Lorentz-factor scaling*.

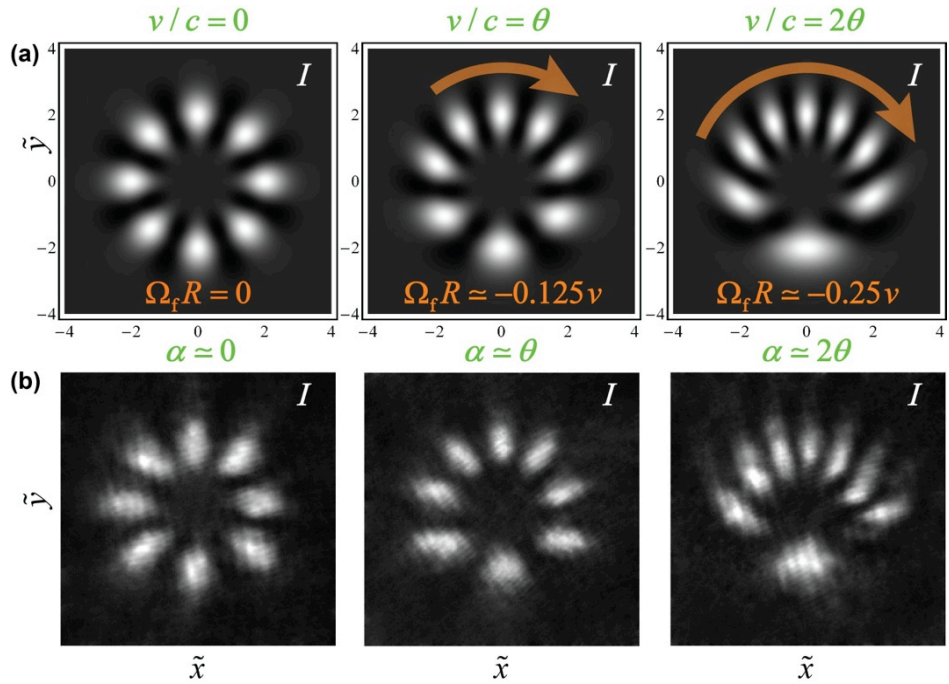

**Figure 4.** (a) Transverse intensity patterns for interference (10) of a vortex beam ($\ell = 8$) and a plane wave propagating along the z-axis in the *moving* reference frame. Thus, the vortex beam and the wave move with respect to each other with non-relativistic velocity $v/c = 0, \theta, 2\theta \ll 1$ in the x-direction. The pronounced y-distortion of the radial fringes visualize the deformed vortex wavefronts shown in Fig. 2, although the fringes do *not* rotate at $v = 0$, when the beam and plane wave have the same frequency $\omega_1 = \omega_2 = \omega$. Alongside the velocity-addition distortions, the fringes start to rotate with the angular velocity (12) and corresponding linear velocity $u_\mathrm{f} \simeq \Omega_\mathrm{f} R \simeq -v^2/8\theta c \sim v$, at $v/c \sim \theta$. This effect originates from the Lorentz-factor scaling of the moving-wave frequency, Eq. (9). (b) Experimental pictures of the interference of an optical vortex beam and a plane wave *tilted* by the angle $\alpha = k_x/k \sim \theta \ll 1$. The precise correspondence between patterns (a) and (b) appears because the Lorentz transformation of time for a transversely moving paraxial wave is equivalent to its tilt (in the approximation $\gamma \simeq 1$).

Recall now that in the problem under consideration, the Lorentz transformation of a paraxial plane wave is equivalent (in the approximation $\gamma \simeq 1$) to the *tilt* of its wave vector by the angle $\alpha \simeq v/c$, Eq. (9). Therefore, the *same y-deformation of the radial interference fringes will appear upon a small x-tilt between the interfering vortex and plane wave*. This effect is familiar to experimentalists working in singular optics. In Figure 4b we show experimentally-measured deformations of the radial interference pattern upon a small tilt between the optical vortex beam and a plane wave. Clearly, Figures 4a and 4b are in perfect agreement with each other. This is explained by the fact that the Lorentz $x$-dependent time delay in Eq. (1) is represented (for waves) by the $x$-dependent *phase* delay, i.e., a tilted wavefront.

Another curious and very close analogy with relativistic velocity-addition deformations occurs in photography, when making pictures of moving objects. Then, the rolling shutter of the camera provides a true $x$-dependent time-delay effect, and blades of a rotating propeller undergo the $y$-distortions shown in Fig. 1b [10,15]. In both of the above analogies, with a tilt and with a rolling shutter, the additional rotation (12) does not occur because it essentially originates from the $\gamma$-factor scaling rather than from the time delay.

## 5. Conclusion

We have considered relativistic deformations of moving objects observed in a moving reference frame. There are two types of such deformations: the Lorentz contraction and distortions arising from the relativistic velocity addition. Considering transverse Lorentz transformations of paraxial waves, we found that the wavefronts experience significant relativistic velocity-addition deformations at non-relativistic velocities (Fig. 2b). This is because of the superluminal phase velocity in the transverse plane. We have shown that such distortions of the wavefronts reveal themselves in the interference with a plane wave moving with respect to the probed wave (Fig. 4a). Furthermore, the Lorentz-factor scaling of the frequency of the moving wave induces an additional motion of the interference fringes. One can say that the *velocity addition* causes *deformation* of the fringes, whereas the Lorentz-factor *deformation* provides *additional velocity* to the fringes.

It should be noticed that an analog of the velocity-addition deformations appear for a small tilt of the plane wave (Fig. 4b). Therefore, to observe a truly relativistic effect, one has to use paraxial waves with a characteristic propagation angle $\theta \ll 1$, a relative transverse motion of the waves with velocity $v \sim \theta c$, while the alignment between the waves should be kept with an accuracy of $|\delta\alpha| < \theta$. One can show that the same conditions are required for the observation of the relativistic motion of the fringes, Eq. (12).

Let us estimate the effects described in this paper for electron vortex beams [12,13]. Taking the reasonable paraxial angle $\theta \sim 10^{-6}$, we find that relativistic effects become noticeable at transverse velocities $v \sim 10^{-6} c \sim 3 \cdot 10^2$ m/s, i.e., at the speed of sound in air. Moreover, for slow massive electrons with momentum $p = \hbar k \ll mc$ and energy $E = \hbar\omega \simeq mc^2$, these conditions can be further relaxed because the transverse phase velocity (4) acquires additional factor $E/pc \gg 1$. Then, the relativistic deformations appear at velocities (5):

$$v \sim \frac{pc}{E}\theta c \ll \theta c .  \qquad (13)$$

This suggests that the relativistic distortions can be observed at velocities $v \sim 1$ cm/s using low-energy ($pc \sim 10$ eV) electron microscopy.

**Acknowledgment**

We are grateful to A.Y. Bekshaev and Y.P. Bliokh for fruitful discussions. This work was supported by the European Commission (Marie Curie Action), Australian Research Council, ARO, JSPS-RFBR contract No. 12-02-92100, Grant-in-Aid for Scientific Research (S), MEXT Kakenhi on Quantum Cybernetics, and the JSPS via its FIRST program.

*Note added.* Because of the time constraints in the publication of the Special Issue, we could not make the second revision requested by an anonymous Referee. As per the Referee's request, here we acknowledge that the superluminal phase velocities are also discussed in the context of Bessel beams and X-shaped waves [5,16]. In addition, we emphasize that the beams shown in Figure 2 in the moving frames represent polychromatic waves. This follows from their transverse motion, which determines the spatio-temporal character of moving vortices [17].